\documentclass[aps,prd,reprint,superscriptaddress]{revtex4-2}
\usepackage{amsmath,amssymb,graphicx,dcolumn, bm}
\usepackage{xcolor}
\usepackage[colorlinks=true,linkcolor=black,citecolor=black,urlcolor=black]{hyperref}
\graphicspath{{figures/}}
\begin{document}
\title{Non-Gaussian Halo Bias from Tensor-Induced Density Perturbations}
\author{Mariam Abdelaziz}
\thanks{mariamtarekmohamed.abdelaziz-ssm@unina.it}
\affiliation{Scuola Superiore Meridionale, Largo San Marcellino 10, I-80138 Napoli, Italy}
\affiliation{INFN, Sezione di Napoli, Via Cinthia Edificio 6, I-80126 Napoli, Italy}
\affiliation{Department of Astronomy, Space Science and Meteorology, Faculty of Science, Cairo University, 12613 Giza, Egypt}

\author{Pritha Bari}
\thanks{prithazxc@gmail.com}
\affiliation{Cosmology, Gravity, and Astroparticle Physics Group,
Center for Theoretical Physics of the Universe,
Institute for Basic Science (IBS), Daejeon 34126, Korea}

\author{Sabino Matarrese}
\thanks{sabino.matarrese@pd.infn.it}
\affiliation{Dipartimento di Fisica e Astronomia ``Galileo Galilei'',
Universit\`a degli Studi di Padova, Via Marzolo 8, I-35131 Padova, Italy}
\affiliation{INFN, Sezione di Padova, Via Marzolo 8, I-35131 Padova, Italy}
\affiliation{INAF, Osservatorio Astronomico di Padova,
Vicolo dell'Osservatorio 5, I-35122 Padova, Italy}
\affiliation{Gran Sasso Science Institute,
Viale F.~Crispi 7, I-67100 L'Aquila, Italy}

\author{Angelo Ricciardone}
\thanks{angelo.ricciardone@unipi.it}
\affiliation{Dipartimento di Fisica ``Enrico Fermi'',
Universit\`a di Pisa, Largo Bruno Pontecorvo 3, I-56127 Pisa, Italy}
\affiliation{INFN, Sezione di Pisa, Largo Bruno Pontecorvo, I-56127 Pisa, Italy}

\begin{abstract}

Primordial gravitational waves induce matter density perturbations at second order. Because this sourcing is quadratic in tensor modes, the induced density field is intrinsically non-Gaussian even for Gaussian tensor modes. Here we compute the scale-dependent halo-bias correction generated by this tensor-induced density bispectrum for a Gaussian-bump tensor spectrum with a peak tensor-to-scalar ratio $r_*=0.002$. At low wavenumbers, both the induced density power spectrum and the bispectrum form factor approach finite, nonzero constants. Their ratio consequently produces a large-scale plateau in the bias correction defined relative to the tensor-induced density power spectrum. Over the sampled scales, the fractional Eulerian correction reaches a maximum of $1.85\times10^{-10}$. When evaluated relative to the total matter power spectrum, the fractional correction remains below $4.0\times10^{-25}$ across the sampled scales and halo populations. Its scale dependence differs from that of the standard local, equilateral, and enfolded primordial non-Gaussianity templates.

\end{abstract}
\maketitle
\section{Introduction}

Primordial gravitational waves (PGWs) could probe the early Universe and, in inflationary scenarios, its characteristic energy scale \cite{Starobinsky:1979ty,Rubakov:1982df,Guth:1980zm,Linde:1981mu,Albrecht:1982wi}.
 Most searches have focused on the $B$-mode polarization of the Cosmic Microwave Background (CMB) \cite{Kamionkowski:2015yta,Guzzetti:2016mkm,LiteBIRD:2022cnt}, interferometric measurements \cite{Maggiore:1999vm,Watanabe:2006qe,Flauger:2020qyi,Caporali:2025mum,Branchesi:2023mws,LISACosmologyWorkingGroup:2024hsc,ET:2025xjr,Ricciardone:2017kre}, and pulsar timing arrays (PTAs) \cite{Campeti:2020xwn,EPTA:2015qep}. PTA collaborations reported evidence for a nanohertz gravitational-wave background in 2023 \cite{NANOGrav:2023gor,EPTA:2023GWB,Xu:2023wog,Reardon:2023GWB}. These measurements have motivated both astrophysical and early-Universe interpretations \cite{NANOGrav:2023hvm,EPTA:2023xxk}.

Large-scale structure (LSS) provides complementary probes of PGWs through tensor fossil signatures \cite{Masui:2010cz,Jeong:2012df,Dai:2013LSS,Dimastrogiovanni:2014ina}, as well as their imprints on galaxy clustering and weak lensing \cite{Kaiser:1996wk,Jeong:2012nu,Schmidt:2012nw}.

Here, we focus on scalar density perturbations induced at second order by primordial tensor modes. Their contribution to the matter density field on sub-horizon scales was studied in \cite{Matarrese:1997ay,Bari:2021xvf,Bari:2022grh}. The same nonlinear tensor-scalar coupling has also been explored as an origin of primordial scalar perturbations in \cite{Bertacca:2024zfb,Traforetti:2025cax,Abdelaziz:2026qyk}.

The quadratic dependence of the induced density perturbation on the tensor field generates intrinsic non-Gaussianity (NG), even when the primordial tensor modes are Gaussian. In our previous work, we computed the induced density bispectrum and showed how its shape depends on the primordial tensor power spectrum \cite{Abdelaziz:2025qpn}. For the Gaussian-bump tensor spectrum considered there, we found
that the induced density bispectrum peaks near equilateral
configurations. Such peaked tensor spectra are motivated by axion-gauge-field inflationary models, including spectator axion-$\mathrm{SU}(2)$ scenarios \cite{Namba:2015gja,Dimastrogiovanni:2016fuu}.

The clustering of dark matter halos responds to the higher-order
correlations of the smoothed density field
\cite{Grinstein:1986en,Matarrese:1986et}. In the high-peak,
large-separation limit, Matarrese and Verde \cite{Matarrese:2008nc}
expressed the leading scale-dependent correction directly in terms
of the density bispectrum. This connection between non-Gaussian density
statistics and halo clustering has been developed and tested for
primordial non-Gaussianity (PNG)
\cite{Dalal:2007cu,Slosar:2008hx,Verde:2009hy,Desjacques:2011mq,Desjacques:2016bnm}.

Here we apply this approach to the tensor-induced density bispectrum
computed in Ref.~\cite{Abdelaziz:2025qpn}. For a Gaussian-bump tensor
spectrum, we calculate the leading halo-bias correction and examine
its dependence on scale, halo mass, and redshift. We relate its
large-scale behavior to the squeezed limit and compare
the scale dependence of the correction defined relative to the
combined scalar and tensor-induced matter power with the local, equilateral, and enfolded PNG templates.


\section{The Tensor-Induced Density Field}
\label{sec:field}

We work with the tensor-induced matter density contrast
$\delta^{(2)}$ in the CDM-comoving synchronous gauge
\cite{Bari:2021xvf,Bari:2022grh,Abdelaziz:2025qpn}. Its matter-era
retarded Green-function solution, with no additional source-free
solution of the density evolution equation, is \cite{Abdelaziz:2025qpn}
\begin{align}
\delta^{(2)}(\mathbf{k},\tau) &= \sum_{\sigma,\sigma'} \int \frac{d^3p}{(2\pi)^3} \, \epsilon^\sigma_{ij}(\hat{\mathbf{p}}) \epsilon^{\sigma' ij}(\widehat{\mathbf{k}-\mathbf{p}}) \nonumber \\
&\quad \times \Bigg[ \frac{\tau^2}{10} \int_0^\tau \frac{d\tilde{\tau}}{\tilde{\tau}} \, \chi'_\sigma(\mathbf{p},\tilde{\tau}) \chi'_{\sigma'}(\mathbf{k}-\mathbf{p},\tilde{\tau}) \nonumber \\
&\quad - \frac{1}{10\tau^3} \int_0^\tau d\tilde{\tau} \, \tilde{\tau}^4 \chi'_\sigma(\mathbf{p},\tilde{\tau}) \chi'_{\sigma'}(\mathbf{k}-\mathbf{p},\tilde{\tau}) \Bigg] ,
\label{eq:kernel}
\end{align}
Here $\tau$ is conformal time and a prime denotes differentiation
with respect to $\tau$. The two tensor polarizations are
$\sigma=+,\times$, with polarization tensors normalized as
$\epsilon_{ij}^{\sigma}(\widehat{\mathbf p})
\epsilon^{\sigma'ij}(\widehat{\mathbf p})
=2\delta_{\sigma\sigma'}$.
For modes entering the horizon during matter domination,
\begin{align}
\chi_\sigma(\mathbf p,\tau)
&=A_\sigma(\mathbf p)\frac{3j_1(p\tau)}{p\tau},
\nonumber\\
\chi_\sigma'(\mathbf p,\tau)
&=-\frac{3A_\sigma(\mathbf p)}{\tau}j_2(p\tau),
\label{eq:densitytransfer}
\end{align}
where $A_\sigma$ are the Gaussian primordial tensor amplitudes
and $j_n$ is the spherical Bessel function of order $n$
\cite{Bari:2021xvf,Abdelaziz:2025qpn}.
The primordial tensor amplitudes satisfy
\begin{align}
\left\langle A_\sigma(\mathbf p)
A_{\sigma'}(\mathbf p')\right\rangle
&=(2\pi)^3\delta_D(\mathbf p+\mathbf p')
\delta_{\sigma\sigma'}
\frac{2\pi^2}{p^3}\Delta_\sigma^2(p),
\end{align}
with $\Delta_\sigma^2(p)=\Delta_T^2(p)/2$, following the
normalization of Ref.~\cite{Abdelaziz:2025qpn}.
Retaining the growing contribution to Eq.~\eqref{eq:kernel} and
substituting Eq.~\eqref{eq:densitytransfer}, we obtain
\cite{Bari:2021xvf,Abdelaziz:2025qpn}
\begin{widetext}
\begin{align}
\delta^{(2)}(\mathbf{k},\tau)
&=\sum_{\sigma,\sigma'}\int\frac{d^3p}{(2\pi)^3}
A_\sigma(\mathbf p)A_{\sigma'}(\mathbf k-\mathbf p)
\epsilon_{ij}^{\sigma}(\hat{\mathbf p})
\epsilon^{\sigma'ij}(\widehat{\mathbf k-\mathbf p})
\left[
\frac{9\tau^2}{10}
\int_0^\tau\frac{d\tilde\tau}{\tilde\tau^3}
j_2(p\tilde\tau)
j_2\!\left(|\mathbf k-\mathbf p|\tilde\tau\right)
\right].
\label{deltaexp}
\end{align}
\end{widetext}

We use the mean-subtracted tensor-induced density field,
\begin{equation}
\delta_T(\mathbf k,\tau)\equiv\delta^{(2)}(\mathbf k,\tau)
 -(2\pi)^3\delta_D(\mathbf k)
 \langle\delta^{(2)}(\mathbf x,\tau)\rangle.
\label{eq:mean_subtracted_field}
\end{equation}
This subtraction removes only the spatial zero mode, $\mathbf k=0$;
the spectra at $k>0$ are unchanged.

Once the tensor source has decayed, the growing solution evolves as
$\delta_T(\mathbf k,z)=D(z)\delta_T(\mathbf k,0)$, with
$D(0)=1$. The corresponding spectra obey

\begin{align}
P_{\delta^{(2)}}(k,z)
&=D^2(z)P_{\delta^{(2)}}(k,0),
\nonumber\\
B_{\delta^{(2)}}(k_i,z)
&=D^3(z)B_{\delta^{(2)}}(k_i,0).
\label{eq:growthstats}
\end{align}
For compactness, we write $P_T\equiv P_{\delta^{(2)}}$ and
$B_T\equiv B_{\delta^{(2)}}$ below. We denote the Gaussian linear
scalar density contrast by $\delta_S\equiv\delta_S^{(1)}$.
Taking the primordial scalar and tensor amplitudes to be statistically
independent \cite{Bari:2021xvf}, the reference matter power spectrum is
\begin{equation}
P_m(k)\equiv P_S(k)+P_T(k),\qquad P_{ST}(k)=0.
\label{eq:reference_matter}
\end{equation}
We calculate the halo-clustering contribution of $B_T$, keeping
scalar nonlinearities and mixed scalar-tensor terms separate
from this tensor-tensor contribution.

\section{Halo Bias Correction}
\label{sec:bias}

In the high-peak threshold model, the smoothed matter density
selects halos according to

\begin{equation}
\rho_{h,R}(\mathbf{x},z_f)
=\Theta\!\left[\delta_R(\mathbf{x})-\delta_c(z_f)\right].
\end{equation}
Here $\delta_R=\delta_{S,R}+\delta_{T,R}$ denotes the smoothed
matter density contrast extrapolated to $z=0$. The smoothing scale
$R$ is associated with halo mass $M$ through
$M=(4\pi/3)\,{\cal A}\,\bar{\rho}_{m,0}\,R^3$, with
$\bar{\rho}_{m,0}=3H_0^2\Omega_{m,0}/(8\pi G)$.
The fudge factor ${\cal A}$ specifies the mass assignment of the
smoothing window. The parameter
$\Delta_c\approx1.68$ is the linearly extrapolated overdensity
for spherical collapse, and $\delta_c(z_f)=\Delta_c/D(z_f)$
defines the corresponding redshift-dependent threshold in terms
of the linear growth factor $D(z)$.
We identify the formation and observation redshifts,
$z_f\simeq z$, in the high-peak approximation
\cite{Matarrese:2008nc}.

For Gaussian initial conditions, this formalism reproduces the standard Kaiser bias \cite{Kaiser:1984sw},
\begin{align}
\label{kaiser}
\xi_{h,M}(r)
&\simeq \exp\!\left[\frac{\nu^2}{\sigma_R^2}\,\xi_R(r)\right]-1 \,,\nonumber\\
&\simeq \frac{\nu^2}{\sigma_R^2}\,\xi_R(r)
= b_{0,L}^2\,\xi_R(r),
\qquad (r\gg R),
\end{align}
where $\nu=\delta_c/\sigma_R$ is the peak height and $\sigma_R$
is the reference root-mean-square density fluctuation extrapolated to $z=0$.
The expansion also requires $b_{0,L}^2|\xi_R(r)|\ll1$.
In the high-peak limit, the Gaussian Lagrangian coefficient is
$b_{0,L}=\delta_c/\sigma_R^2$ \cite{Matarrese:2008nc}.
The Press-Schechter mass-function result,
$b_L^{\rm PS}=(\nu^2-1)/\delta_c$
\cite{Mo:1995cs,Catelan:1997qw}, has the same leading term and
an additional contribution $-1/\delta_c$.

For the numerical calculation, we evaluate the reference variance
with a top-hat window (${\cal A}=1$),
$\sigma_R=\sigma_{\rm TH}(R_{\rm TH})$, and smooth the tensor-induced
density contrast entering the bispectrum with a Gaussian window
(${\cal A}=3.77$) at $R_G=R_{\rm TH}/3.77^{1/3}$.
Both radii correspond to the same halo mass.
Appendix~\ref{app:response} quantifies the normalization dependence
on the reference variance.

For general non-Gaussian initial conditions, the halo correlation function receives higher-order contributions. Following~\cite{Matarrese:1986et,Matarrese:2008nc,Grinstein:1986en} and retaining only the leading non-Gaussian term, $N=3$, one obtains
\begin{widetext}
\begin{align}
\label{eq:xi_expansion}
\xi_{h,M}(|\mathbf{x}_1-\mathbf{x}_2|)
&\simeq -1+\exp\left(
\sum_{N=2}^{\infty}\frac{\nu^N}{\sigma_R^N}
\sum_{j=1}^{N-1}\frac{1}{j!(N-j)!}
\,\xi^{(N)}\left(
\underbrace{\mathbf{x}_1,\ldots,\mathbf{x}_1}_{j\text{ times}},
\underbrace{\mathbf{x}_2,\ldots,\mathbf{x}_2}_{(N-j)\text{ times}}
\right)\right)\,,\nonumber\\
&\simeq b_{0,L}^2\,\xi^{(2)}(\mathbf{x}_1,\mathbf{x}_2)
+b_{0,L}^3\,\xi^{(3)}(\mathbf{x}_1,\mathbf{x}_1,\mathbf{x}_2),
\end{align}
\end{widetext}
where $\xi^{(N)}$ denotes the connected $N$-point correlation function of the underlying smoothed density field.
We now isolate the contribution of the tensor-induced density
field to the halo correlation function. A long-wavelength tensor-induced
density mode shifts the collapse threshold in the same way as a scalar
mode; the Gaussian reference abundance thus responds through the
coefficients $b_{1,L}$ and $b_{2,L}$ defined in
Appendix~\ref{app:response}. Recalling the two-point function and leading connected three-point term gives
\begin{align}
\xi_{h^{(2)},M}(|\mathbf{x}_1-\mathbf{x}_2|)
&\simeq b_{1,L}^2\,\xi_{\delta_T}^{(2)}
(\mathbf{x}_1,\mathbf{x}_2)\nonumber\\
&\quad+b_{1,L}b_{2,L}\,\xi_{\delta_T}^{(3)}
(\mathbf{x}_1,\mathbf{x}_1,\mathbf{x}_2).
\end{align}

At high peak height, Appendix~\ref{app:response} gives
$b_{1,L}\simeq b_{0,L}=\delta_c/\sigma_R^2$ and
$b_{2,L}\simeq b_{0,L}^2$, so that the bispectrum coefficient
reduces to $b_{0,L}^3$.

On large scales, $kR_G\ll1$, Fourier transforming the terms sourced by $\delta^{(2)}$ gives
\begin{widetext}
\begin{align}
P_{h^{(2)}}(k)
&\simeq b_{0,L}^2\,P_T(k)
+\frac{b_{0,L}^3}{(2\pi)^3}
\int d^3k_1\,
W_{R_G}(k_1)W_{R_G}(|\mathbf{k}+\mathbf{k}_1|)W_{R_G}(k)
B_T(k_1,|\mathbf{k}+\mathbf{k}_1|,k)\nonumber\\
&\equiv b_{0,L}^2\,P_T(k)
+\Delta P_{h^{(2)}}(k),
\label{Main1}
\end{align}
\end{widetext}
where $P_T$ and $B_T$ are the induced
density power spectrum and bispectrum extrapolated to $z=0$
\cite{Abdelaziz:2025qpn}. The notation $P_{h^{(2)}}$ identifies
the tensor-induced terms in the halo power, not a separate
population of halos.

With $W_{R_G}(k)=\exp[-(kR_G)^2/2]$, the bispectrum contribution
becomes
\begin{align}
\Delta P_{h^{(2)}}(k)
&=\frac{b_{0,L}^3}{(2\pi)^2}
\int_0^\infty k_1^2\,dk_1\nonumber\\
&\quad\times\int_{-1}^{1}d\mu\,
e^{-R_G^2(k_1^2+k^2+kk_1\mu)}
B_T(k_1,k',k),
\label{deltaP}
\end{align}
where $\mu=\widehat{\mathbf{k}}_1\cdot\widehat{\mathbf{k}}$ and $k'=\sqrt{k_1^2+k^2+2kk_1\mu}$.

Following~\cite{Matarrese:2008nc}, Eq.~\eqref{Main1} can be recast in terms of a scale-dependent bias modulation,
\begin{align}
P_{h^{(2)}}(k)
= b_{0,L}^2\,P_T(k)
\left[
1+\delta_c\,e^{-(kR_G)^2}\,
\frac{\mathcal{F}^{(2)}(k)}{P_T(k)}
\right],
\end{align}
where the form factor is
\begin{align}
\mathcal{F}^{(2)}(k)
&=\frac{1}{(2\pi\sigma_R)^2}
\int_0^\infty k_1^2\,dk_1\nonumber\\
&\quad\times\int_{-1}^{1}d\mu\,
e^{-R_G^2(k_1^2+kk_1\mu)}
B_T(k_1,k',k).
\end{align}

Normalizing the bispectrum contribution to $P_T$ gives the
effective Lagrangian bias correction
\begin{align}
\Delta b_L^{(T)}(k)
&=\frac{b_{0,L}\delta_c}{2}\,
 e^{-(kR_G)^2}\frac{\mathcal F^{(2)}(k)}{P_T(k)}.
\label{lagrangianbias}
\end{align}
Here $b_{0,L}$ and $\Delta b_L^{(T)}(k)$ carry an
implicit redshift dependence through
$\delta_c(z)=\Delta_c/D(z)$.
To first order in $B_T$, this gives
$P_{h^{(2)}}\simeq[b_{0,L}+\Delta b_L^{(T)}]^2P_T$.
The Lagrangian coefficients above multiply the density field
extrapolated to $z=0$. For halos comoving with matter, the
large-scale Eulerian conversion is therefore
\begin{equation}
b_{0,E}(z)=1+\frac{b_{0,L}}{D(z)},\qquad
\Delta b_E^{(T)}(k,z)=\frac{\Delta b_L^{(T)}(k)}{D(z)}.
\label{eq:eulerian_conversion}
\end{equation}
The fractional correction is
\begin{equation}
\frac{\Delta b_E^{(T)}(k,z)}{b_{0,E}(z)}
=\frac{\Delta b_L^{(T)}(k)}{D(z)+b_{0,L}}.
\label{eq:frac_tensor_component}
\end{equation}

Including the Gaussian linear scalar contribution, and retrieving
the two-point terms and the leading term in $B_T$, gives
\begin{align}
P_{hh}^{L}(k)
&=b_{0,L}^2[P_S(k)+P_T(k)]
+\Delta P_{h^{(2)}}(k),
\nonumber\\
\Delta P_{h^{(2)}}(k)
&=b_{0,L}^3\sigma_R^2e^{-(kR_G)^2}\mathcal F^{(2)}(k).
\label{eq:full_halo_power}
\end{align}
The contribution of the same bispectrum $B_T$ to the bias
defined relative to $P_m=P_S+P_T$ is
\begin{align}
\Delta b_L^{\mathrm{tot}}(k)
&=\frac{b_{0,L}\delta_c}{2}e^{-(kR_G)^2}
\frac{\mathcal F^{(2)}(k)}{P_m(k)}
\nonumber\\
&=\frac{P_T(k)}{P_m(k)}\Delta b_L^{(T)}(k),
\label{eq:total_bias}
\end{align}
and
\begin{equation}
\frac{\Delta b_E^{\mathrm{tot}}(k,z)}{b_{0,E}(z)}
=\frac{\Delta b_L^{\mathrm{tot}}(k)}{D(z)+b_{0,L}}.
\label{obsbias}
\end{equation}
Appendix~\ref{app:response} gives the corresponding halo-matter
cross-spectrum derivation, and Appendix~\ref{app:diagnostics} presents
the absolute and signed bias corrections.


\section{Results}
We evaluate the tensor-induced halo-bias correction in a flat
$\Lambda$CDM cosmology with Planck 2018 parameters
$(\Omega_m,h)=(0.31,0.67)$, $A_s=2.1\times10^{-9}$, and
$n_s=0.965$ at the pivot scale $k_0=0.05\,\mathrm{Mpc}^{-1}$
\cite{Planck:2018vyg}.

\subsection{Tensor-Spectrum Normalization}
We consider a primordial tensor spectrum characterized by a Gaussian bump,
\begin{equation}
\Delta_T^2(k)=A_T\exp\left[-\frac{\ln^2(k/k_p)}{2\sigma^2}\right],
\label{eq:bump}
\end{equation}
centered at $k_p=0.002\,\mathrm{Mpc}^{-1}$ with width $\sigma=0.70$.
Setting the peak tensor-to-scalar ratio to $r_*=0.002$ gives
$A_T=r_*\Delta_\zeta^2(k_p)$, where
$\Delta_\zeta^2(k)=A_s(k/k_0)^{n_s-1}$, giving
\begin{equation}
A_T= 4.70086\times10^{-12}.
\label{eq:benchmark}
\end{equation}
For a fixed bump shape, $P_T\propto A_T^2$ and $B_T\propto A_T^3$,
so that $\Delta b_L^{(T)}\propto A_T$, whereas
$\Delta b_L^{\mathrm{tot}}\propto A_T^3$ when $P_S\gg P_T$.

\subsection{Mass, Redshift, and Scale Dependence}

\begin{figure*}[t]
\centering
\includegraphics[width=\textwidth]{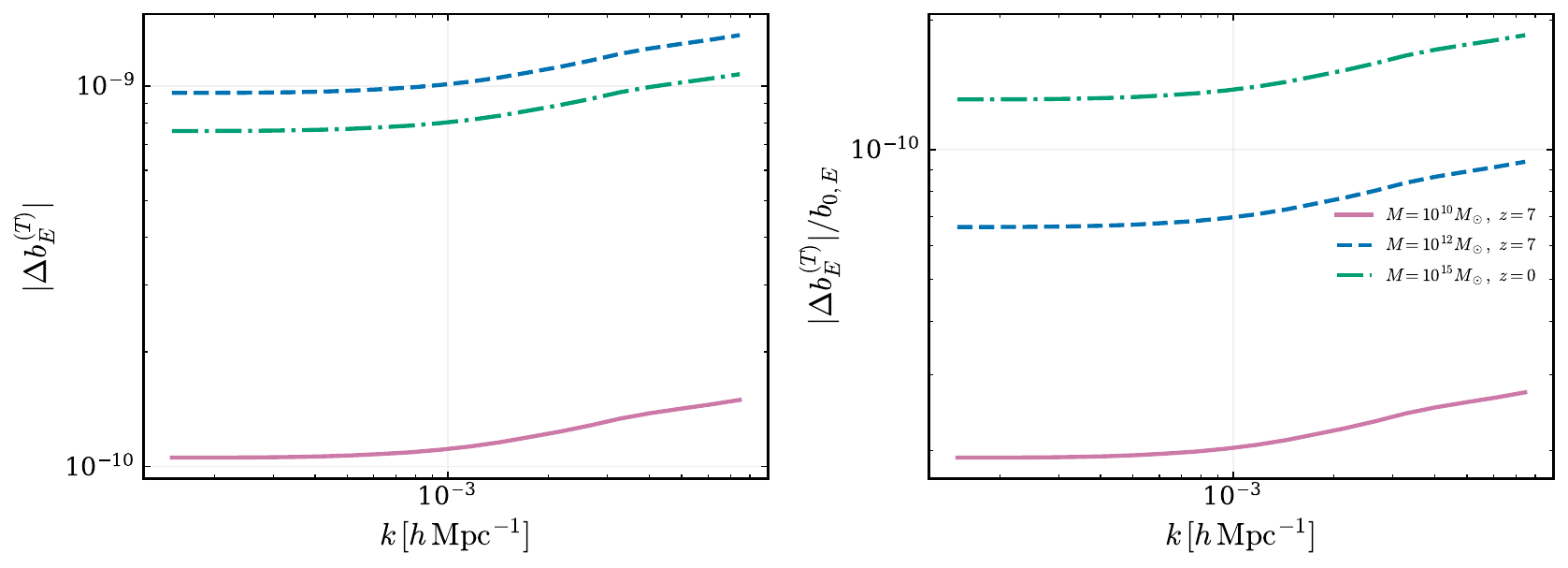}
\caption{Tensor-component Eulerian bias correction for the
Gaussian-bump spectrum with $r_*=0.002$ at
$k_p=0.002\,\mathrm{Mpc}^{-1}$ and $\sigma=0.70$.
Left: $|\Delta b_E^{(T)}(k,z)|$; right:
$|\Delta b_E^{(T)}|/b_{0,E}$. The three samples are
$M=10^{10},10^{12}\,M_\odot$ at $z=7$ and
$M=10^{15}\,M_\odot$ at $z=0$. The correction is normalized to the tensor-induced density power.}
\label{fig:biasfrac}
\end{figure*}

Figure~\ref{fig:biasfrac} presents the absolute and fractional
Eulerian bias corrections normalized to the tensor-induced component,
for $1.49\times10^{-4}\leq k\leq7.46\times10^{-3}\,
 h\,\mathrm{Mpc}^{-1}$ (equivalently, $10^{-4}\leq k\leq5\times10^{-3}\,\mathrm{Mpc}^{-1}$ for $h=0.67$). We consider halos of
$M=10^{10}$ and $10^{12}\,M_\odot$ at $z=7$ and
$M=10^{15}\,M_\odot$ at $z=0$.

The corrections are positive over the sampled range, so the
absolute values displayed in Fig.~\ref{fig:biasfrac} coincide with
the signed corrections. The largest fractional correction is $1.85\times10^{-10}$ for
$M=10^{15}\,M_\odot$ at $z=0$, while the
$M=10^{12}\,M_\odot$ sample at $z=7$ reaches
$9.38\times10^{-11}$. The curves approach a plateau at low $k$ and
rise gradually toward higher wavenumbers.

\begin{figure*}[t]
\centering
\includegraphics[width=\textwidth]{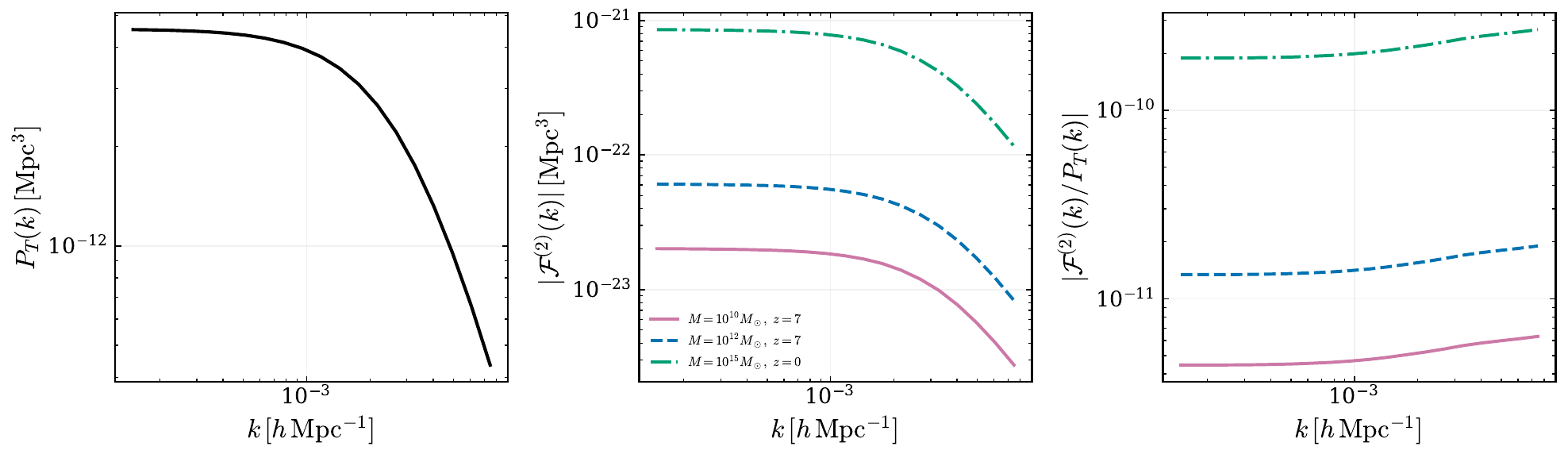}
\caption{The tensor-induced density power spectrum and bispectrum form
factor entering the halo-bias correction.
Left: $P_T(k)$; middle: the smoothed form factor
$|\mathcal F^{(2)}(k)|$; right:
$|\mathcal F^{(2)}(k)/P_T(k)|$.
The form factor is shown for the three halo samples.
All spectra are extrapolated to $z=0$.
The low-$k$ behavior is derived in Appendix~\ref{app:squeezed}.}
\label{fig:lowk}
\end{figure*}

For $k\ll k_1$, the form factor samples squeezed configurations,
$B_T(k_1,|\mathbf k_1+\mathbf k|,k)$. Appendix~\ref{app:squeezed}
shows that the squeezed bispectrum and $P_T(k)$ have finite low-$k$
limits. The form factor and $P_T(k)$ approach positive constants, so
$\mathcal F^{(2)}(k)/P_T(k)$ also approaches a constant (see Fig.~\ref{fig:lowk}) and gives
the plateau in Fig.~\ref{fig:biasfrac}.

\subsection{Comparison with Standard PNG Templates}

The total-matter response instead involves $\mathcal F^{(2)}(k)/P_m(k)$
and shows a different scale dependence. We compare the fractional
total-matter correction,
$|\Delta b_E^{\mathrm{tot}}|/b_{0,E}$, with the local, equilateral,
and enfolded PNG templates \cite{Verde:2009hy}. To compare their
scale dependence, we rescale each PNG template to match the mean
tensor-induced total-matter correction over the three highest
sampled wavenumbers.

\begin{figure*}[t]
\centering
\includegraphics[width=\textwidth]{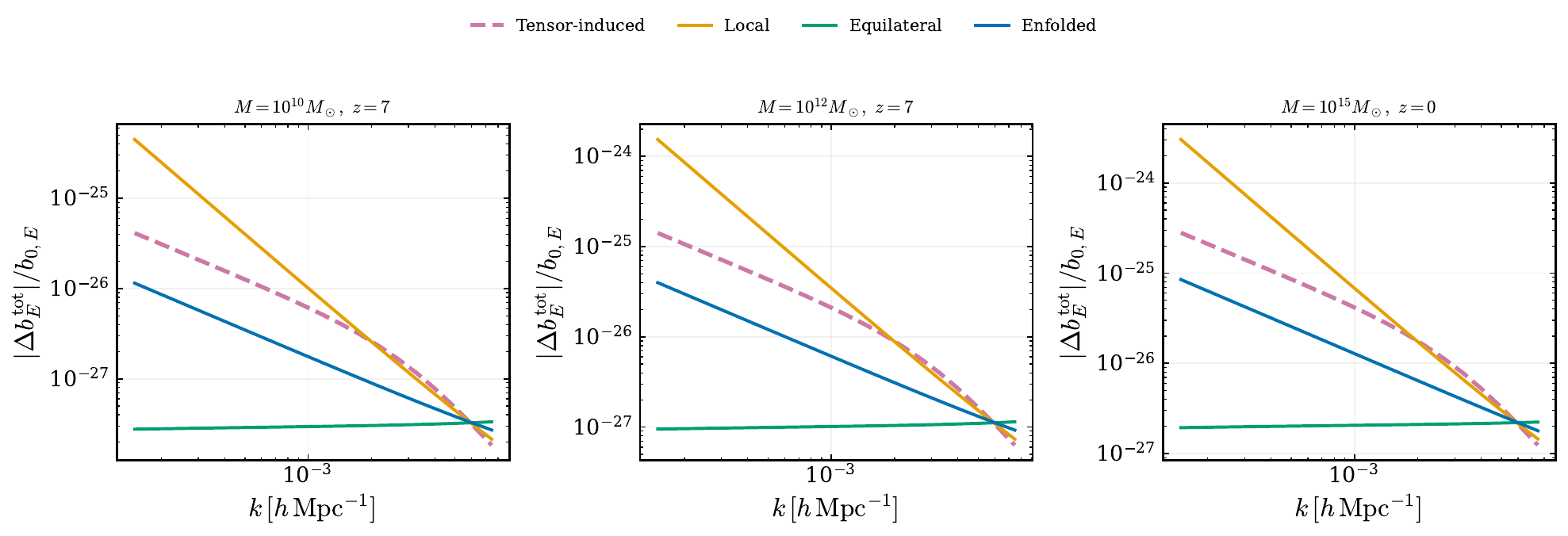}
\caption{Scale dependence of the tensor-induced fractional Eulerian
bias correction normalized to the total matter power (dashed),
compared with the local, equilateral, and enfolded PNG templates
(solid). The three halo samples are
$M=10^{10},10^{12}\,M_\odot$ at $z=7$ and
$M=10^{15}\,M_\odot$ at $z=0$.
Each PNG template is rescaled to match the mean tensor-induced
total-matter correction over the three highest sampled wavenumbers.}
\label{fig:templatescompare}
\end{figure*}

Figure~\ref{fig:templatescompare} shows that the tensor-induced
correction differs in scale dependence from each individual template.
On scales where the scalar matter transfer function approaches unity,
$P_S(k)\propto k^{n_s}$. With a constant low-$k$ form factor and
$P_S\gg P_T$, this gives
\begin{equation}
\Delta b_L^{\mathrm{tot}}(k)\propto P_S^{-1}(k)
\propto k^{-n_s},
\label{eq:scalar_dominated_scaling}
\end{equation}
rather than the plateau of the correction normalized to $P_T$.
Appendix~\ref{app:diagnostics} gives the total-matter bias amplitude
and the corresponding halo-power contributions.

\section{Conclusions}

Primordial gravitational waves can influence cosmic structure beyond
their contribution to the matter power spectrum. By sourcing density
perturbations at second order, they generate an induced density field
that is intrinsically non-Gaussian, even when the primordial tensor
fluctuations are Gaussian. In this work, we have investigated how
this non-Gaussianity enters the large-scale clustering of dark matter
halos, extending our previous analysis of the tensor-induced density
bispectrum to its consequences for halo bias.

Using the high-peak, large-separation formalism of Matarrese and
Verde, we derived the leading halo-bias correction sourced by the
tensor-induced density bispectrum and examined its dependence on
scale, halo mass, and redshift. For the peaked tensor spectrum
studied here, the bispectrum and the induced density power
approach finite, nonzero limits at low wavenumber. The resulting
response, when normalized to the tensor-induced density power,
therefore approaches a plateau on large scales and rises gradually
toward higher wavenumbers over the sampled range. This behavior
connects the large-scale halo response to the mode coupling generated
by the quadratic dependence of the induced density field on the
primordial tensor fluctuations.

For a Gaussian-bump tensor spectrum centered at
$k_p=0.002\,\mathrm{Mpc}^{-1}$, with width $\sigma=0.70$ and peak
tensor-to-scalar ratio $r_*=0.002$, the largest tensor-component
fractional Eulerian correction is $1.85\times10^{-10}$, attained
for $M=10^{15}\,M_\odot$ halos at $z=0$ among the samples considered.
When evaluated relative to the combined linear and tensor-induced
matter power, the corresponding fractional correction is bounded
by $3.96\times10^{-25}$ over the sampled wavenumber interval. The scale dependence of this total-matter
correction differs from each of the local, equilateral, and enfolded
primordial non-Gaussianity templates considered individually.

Our results connect the non-Gaussian statistics of tensor-induced
density perturbations to the clustering of dark matter halos.
Halo bias thus carries information about tensor-induced mode
coupling that is not contained in the induced matter power spectrum
alone. Applying the same calculation to other primordial tensor
spectra would reveal how their different density bispectra translate
into halo-clustering signatures, complementing the study of their
higher-order correlations in the matter distribution.
\begin{acknowledgments}
The authors thank Giulia Rodighiero for useful discussions and helpful input. MA thanks Licia Verde and the cosmology group at the Institute of Cosmos Sciences of the University of Barcelona (ICCUB) for useful discussions. PB is supported by IBS under the project code IBS-R018-D3. SM acknowledges partial financial support from the COSMOS network (www.cosmosnet.it) through ASI Grants 2016-24-H.0, 2016-24-H.1-2018 and 2020-9-HH.0.
\end{acknowledgments}

\appendix
\section{Squeezed limit of the tensor-induced density statistics}
\label{app:squeezed}

\begin{figure*}[t]
\centering
\includegraphics[width=\textwidth]{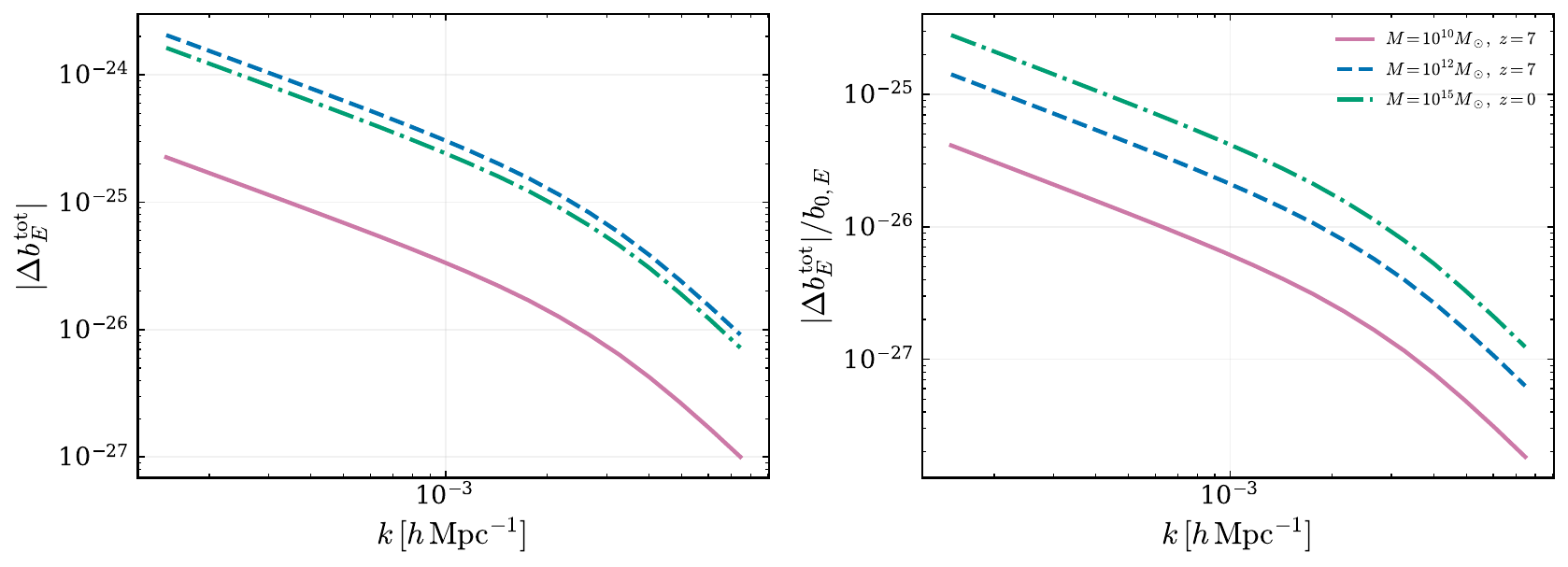}
\caption{Total-matter Eulerian bias correction and
fractional correction for the three halo samples, computed
with $P_m=P_S+P_T$ in Eq.~\eqref{eq:total_bias}.}
\label{fig:app_total_bias}
\end{figure*}

\begin{figure*}[t]
\centering
\includegraphics[width=\textwidth]{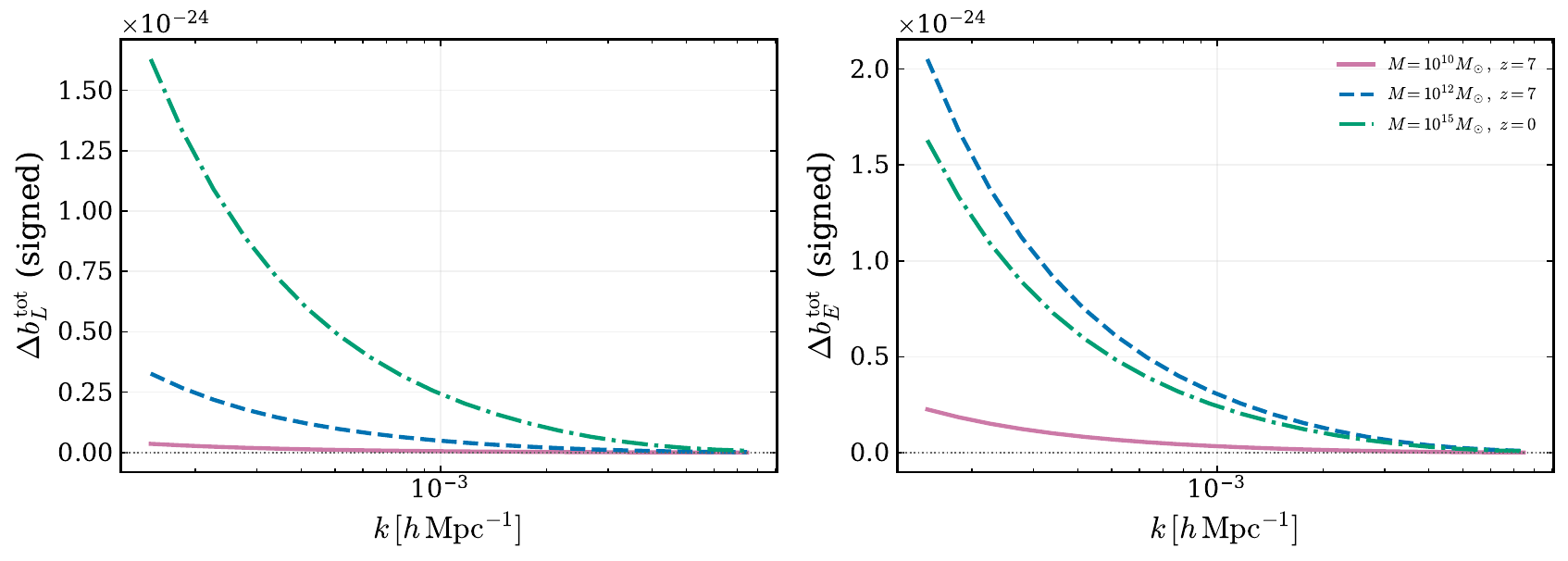}
\caption{Signed total-matter Lagrangian
$\Delta b_L^{\mathrm{tot}}(k)$ and Eulerian
$\Delta b_E^{\mathrm{tot}}(k,z)$ bias corrections for the three
halo samples. The corrections are positive throughout the sampled
wavenumber interval.}
\label{fig:app_signed}
\end{figure*}

The tensor-induced density field takes the quadratic form
\begin{align}
\delta_T(\mathbf k,\tau)
&=\int\frac{d^3p}{(2\pi)^3}\,
\mathcal K_\tau(p,q)\,
X_{ij}(\mathbf p)X_{ij}(\mathbf k-\mathbf p),
\nonumber\\
\mathcal K_\tau(p,q)
&=\frac{9\tau^2}{10}\int_0^\tau
\frac{d\tilde\tau}{\tilde\tau^3}
 j_2(p\tilde\tau)j_2(q\tilde\tau),
\label{eq:app_kernel}
\end{align}
where $q=|\mathbf k-\mathbf p|$, $k>0$, and
$X_{ij}(\mathbf p)=\sum_\sigma
\epsilon^\sigma_{ij}(\hat{\mathbf p})A_\sigma(\mathbf p)$.
The covariance of $X_{ij}$ is
\begin{align}
\langle X_{ij}(\mathbf p) X_{mn}(\mathbf p')\rangle
&=(2\pi)^3\delta_D(\mathbf p+\mathbf p')
\mathcal P_A(p)\Pi_{ij,mn}(\hat{\mathbf p}),
\nonumber\\
\mathcal P_A(p)&=\frac{2\pi^2}{p^3}\frac{\Delta_T^2(p)}{2},
\label{eq:app_covariance}
\end{align}
with the polarization projector
\begin{align}
\Pi_{ij,mn}(\hat{\mathbf p})
&=P_{im}P_{jn}+P_{in}P_{jm}-P_{ij}P_{mn},
\nonumber\\
P_{ij}&=\delta_{ij}-\hat p_i\hat p_j.
\label{eq:app_projector}
\end{align}
The normalization $\epsilon^\sigma_{ij}\epsilon^{\sigma'ij}
=2\delta_{\sigma\sigma'}$ implies
$\Pi^2=2\Pi$ when viewed as a matrix on pairs of indices.

For Gaussian primordial tensor amplitudes, the connected density
bispectrum is the sum of eight triangular Wick contractions. With
$\mathbf k_1+\mathbf k_2+\mathbf k_3=0$,
$\mathbf q=\mathbf k_1-\mathbf p$, and
$\mathbf r=\mathbf k_2+\mathbf p$, the result is

\begin{widetext}

\begin{align}
B_T(k_1,k_2,k_3)
=8\int\frac{d^3p}{(2\pi)^3}\,
&\mathcal P_A(p)\mathcal P_A(q)\mathcal P_A(r)
\mathcal K_\tau(p,q)\mathcal K_\tau(p,r)
\mathcal K_\tau(q,r)
\Pi_{ij,mn}(\hat{\mathbf p})
\Pi_{mn,ab}(\hat{\mathbf q})
\Pi_{ab,ij}(\hat{\mathbf r}).
\label{eq:app_bispectrum_wick}
\end{align}

\end{widetext}

In the squeezed configuration $\mathbf k_2\to-\mathbf k_1$ and
$k_3\to0^+$, one has $\mathbf r\to-\mathbf q$. With
$\mu=\hat{\mathbf p}\cdot\hat{\mathbf q}$, the polarization contraction is
\begin{align}
&\Pi_{ij,mn}(\hat{\mathbf p})
\Pi_{mn,ab}(\hat{\mathbf q})
\Pi_{ab,ij}(\hat{\mathbf q})=2\left(1+6\mu^2+\mu^4\right).
\label{eq:app_polarization_identity}
\end{align}
This follows from Eq.~\eqref{eq:app_projector}.
The squeezed limit of Eq.~\eqref{eq:app_bispectrum_wick} is therefore

\begin{widetext}

\begin{equation}
\lim_{k_L\to0^+}B_T(k_s,k_s,k_L)
=16\int\frac{d^3p}{(2\pi)^3}\,
\mathcal P_A(p)\mathcal P_A^2(q)
\mathcal K_\tau^2(p,q)\mathcal K_\tau(q,q)
(1+6\mu^2+\mu^4),\qquad
q=|\mathbf k_s-\mathbf p|.
\label{eq:app_squeezed_bispectrum}
\end{equation}

\end{widetext}

The integrand is nonnegative, and the localized tensor spectrum
gives a finite, positive squeezed bispectrum. The induced power spectrum
likewise satisfies
\begin{equation}
\lim_{k\to0^+}P_T(k)
=16\int\frac{d^3p}{(2\pi)^3}\,
\mathcal P_A^2(p)\mathcal K_\tau^2(p,p)>0.
\label{eq:app_small_k_power}
\end{equation}
Defining $B_T^{\rm sq}(k_s)\equiv\lim_{k_L\to0^+}
B_T(k_s,k_s,k_L)$, the form factor satisfies
\begin{align}
\lim_{k\to0^+}\mathcal F^{(2)}(k)
&=\frac{2}{(2\pi\sigma_R)^2}
\int_0^\infty dk_s\,k_s^2 e^{-R_G^2k_s^2}\nonumber\\
&\qquad\times B_T^{\rm sq}(k_s)>0.
\label{eq:app_small_k_F}
\end{align}
Both quantities tend to finite constants, giving the plateau of
$\mathcal F^{(2)}(k)/P_T(k)$ in Fig.~\ref{fig:lowk}.

\section{Threshold response and halo-matter cross-bias}
\label{app:response}

The collapse condition acts on
$\delta_R=\delta_{S,R}+\delta_{T,R}$. A long mode
$\Delta_L=\delta_{S,L}+\delta_{T,L}$ shifts the threshold to
$\delta_c-\Delta_L$. For the Gaussian reference abundance, with
$\nu=\delta_c/\sigma_R$ \cite{Kaiser:1984sw,Matarrese:2008nc,Slosar:2008hx},
\begin{align}
\bar n_h(\Delta_L)&\propto
Q\!\left(\nu-\frac{\Delta_L}{\sigma_R}\right),\nonumber\\
Q(\nu)&=\frac12\operatorname{erfc}\!\left(\frac{\nu}{\sqrt2}\right).
\label{eq:app_abundance}
\end{align}
The response coefficients, defined by
$b_{n,L}=\bar n_h(0)^{-1}\partial_{\Delta_L}^n
\bar n_h|_{\Delta_L=0}$, are
\begin{align}
b_{1,L}&=\frac{\varphi(\nu)}{\sigma_R Q(\nu)},
\nonumber\\
b_{2,L}&=\frac{\nu\varphi(\nu)}{\sigma_R^2Q(\nu)},
\qquad
\varphi(\nu)=\frac{e^{-\nu^2/2}}{\sqrt{2\pi}}.
\label{eq:app_response_coefficients}
\end{align}
Since the abundance depends on the sum of the long modes, their
linear density responses are equal:
\begin{equation}
\frac{\partial\ln\bar n_h}{\partial\delta_{S,L}}
=\frac{\partial\ln\bar n_h}{\partial\delta_{T,L}}
=b_{1,L}.
\label{eq:app_equal_response}
\end{equation}
At $\nu\gg1$, the asymptotic form
$Q(\nu)\sim\varphi(\nu)/\nu$ yields
\begin{equation}
b_{1,L}\simeq\frac{\delta_c}{\sigma_R^2}
\equiv b_{0,L},\qquad b_{2,L}\simeq b_{0,L}^2.
\label{eq:app_highpeak_coeffs}
\end{equation}
Writing the smoothed bispectrum contribution as
\begin{equation}
\mathcal I_T(k)\equiv
\sigma_R^2e^{-(kR_G)^2}\mathcal F^{(2)}(k),
\end{equation}
its coefficient in the halo power is
$\Delta P_{hh}^{L}=b_{1,L}b_{2,L}\mathcal I_T$,
which becomes the $b_{0,L}^3$ term of
Eq.~\eqref{eq:full_halo_power} in the high-peak limit.
The halo-matter cross-spectrum gives the same correction at
leading order in $kR_G$. We write $W_G(kR_G)\equiv W_{R_G}(k)$ for the Gaussian window defined in Sec.~\ref{sec:bias}.
The Gaussian-weighted convolution is
\begin{align}
\mathcal J_T(k) &\equiv \int\frac{d^3q}{(2\pi)^3}
W_G(qR_G)W_G(|\mathbf k-\mathbf q|R_G)\nonumber\\
&\qquad\times B_T(q,|\mathbf k-\mathbf q|,k).
\label{eq:app_J_cross}
\end{align}
At leading bispectrum order, the halo-matter cross-spectrum is
\begin{equation}
P_{hm}^{L}(k)=b_{1,L}P_m(k)
+\frac{b_{2,L}}{2}\mathcal J_T(k)
+\mathcal O[(kR_{\rm TH})^2P_m(k)].
\label{eq:app_cross_spectrum}
\end{equation}
The corresponding cross-bias correction is
\begin{align}
\Delta b_{L,hm}(k)
&=\frac{b_{2,L}}{2}\frac{\mathcal J_T(k)}{P_m(k)}
\nonumber\\
&=\frac{b_{2,L}\sigma_R^2}{2}
 e^{-(kR_G)^2/2}\frac{\mathcal F^{(2)}(k)}{P_m(k)}.
\label{eq:app_cross_bias}
\end{align}
In the high-peak limit, $b_{2,L}\sigma_R^2\simeq b_{0,L}\delta_c$.
The auto-spectrum expression retains one additional external window,
so that, at the retained bispectrum order,
\begin{equation}
\Delta b_L^{\mathrm{tot}}(k)
\simeq W_G(kR_G)\,\Delta b_{L,hm}(k).
\label{eq:app_auto_cross_window}
\end{equation}
Thus the two definitions agree at leading order in $kR_G$.
For the largest smoothing radius in our sample,
$R_G=11.792\,\mathrm{Mpc}$, and
$k\leq5\times10^{-3}\,\mathrm{Mpc}^{-1}$,
$|1-W_G(kR_G)|<1.8\times10^{-3}$.
The Eulerian conversion used in the main text is understood at
the same order in the large-scale expansion.

The two smoothing windows are
\begin{align}
W_{\rm TH}(x)&=3\frac{\sin x-x\cos x}{x^3},
\qquad W_G(x)=e^{-x^2/2},\nonumber\\
\sigma_X^2(R_X)&=\int d\ln q\,\frac{q^3P_S(q,0)}{2\pi^2}
 W_X^2(qR_X),\quad X={\rm TH},G.
\label{eq:app_windows}
\end{align}
We evaluate the halo coefficients with $\sigma_R=\sigma_{\rm TH}$
and the bispectrum convolution with $W_G$. Holding the Gaussian
window inside the bispectrum integral fixed, replacing the top-hat
reference variance $\sigma_{\rm TH}^2$ by $\sigma_G^2$ changes
$|\Delta b_E^{\mathrm{tot}}|/b_{0,E}$ by factors
$1.26$, $1.34$, and $1.76$ for
$(M,z)=(10^{10}M_\odot,7)$, $(10^{12}M_\odot,7)$, and
$(10^{15}M_\odot,0)$, respectively. The change is in normalization
at fixed halo mass and redshift.

\section{Total-matter bias and halo-power corrections}
\label{app:diagnostics}

\begin{figure*}[t]
\centering
\includegraphics[width=\textwidth]{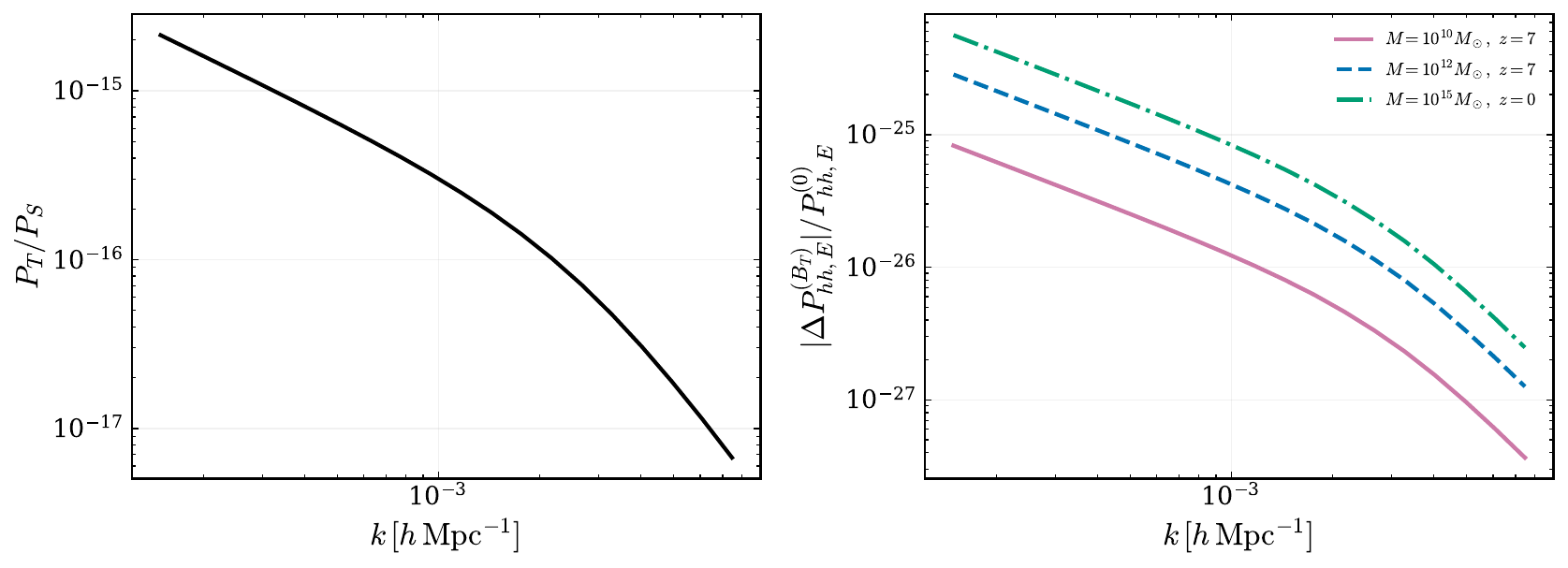}
\caption{Tensor-induced contributions to matter and halo
clustering: $P_T/P_S$ and the bispectrum-sourced fractional change
$\Delta P_{hh,E}^{(B_T)}/P_{hh,E}^{(0)}$.}
\label{fig:app_power}
\end{figure*}

Over the sampled wavenumber interval, the adopted spectrum gives
\begin{equation}
\max_k\frac{P_T(k)}{P_S(k)}\simeq2.14\times10^{-15}.
\label{eq:app_power_ratio}
\end{equation}
For the same interval and halo samples, the total-matter
correction satisfies
\begin{align}
\frac{\Delta b_E^{\mathrm{tot}}}{b_{0,E}}
&=\frac{P_T}{P_m}\frac{\Delta b_E^{(T)}}{b_{0,E}},
\nonumber\\
\max_{k,M,z}\left|\frac{\Delta b_E^{\mathrm{tot}}}{b_{0,E}}\right|
&\leq3.96\times10^{-25}.
\label{eq:app_total_bound}
\end{align}
The upper bound is obtained by multiplying the maximum power
ratio in Eq.~\eqref{eq:app_power_ratio} by the largest
fractional Eulerian correction normalized to $P_T$. These maxima
need not occur at the same wavenumber.

At leading order in the large-scale expansion, the Eulerian
halo power and its bispectrum correction are
\begin{align}
P_{hh,E}^{(0)}(k,z)&=b_{0,E}^2(z)D^2(z)P_m(k,0),
\nonumber\\
\Delta P_{hh,E}^{(B_T)}(k,z)
&=2b_{0,E}(z)\Delta b_E^{\mathrm{tot}}(k,z)
  D^2(z)P_m(k,0).
\label{eq:app_halo_power}
\end{align}
At two-point order, the induced-to-scalar halo-power ratio is
\begin{equation}
\frac{P_{hh,E}^{(T,2\mathrm{pt})}}{P_{hh,E}^{(S,2\mathrm{pt})}}
=\frac{P_T}{P_S}.
\label{eq:app_two_point_fraction}
\end{equation}
Figures~\ref{fig:app_total_bias}-\ref{fig:app_power} show the
total and signed bias corrections and the power-spectrum ratios. The
signed corrections are positive over the sampled interval.

\bibliographystyle{apsrev4-2}
\bibliography{ref}
\end{document}